\documentclass[aps,pra,twocolumn,amsmath,amssymb,nofootinbib,superscriptaddress]{revtex4-2}
\usepackage{times}
\usepackage[pdftex]{graphicx}
\usepackage{dcolumn}
\usepackage{bm}
\usepackage{amsmath}
\usepackage{amsthm}
\usepackage{braket}
\usepackage{indentfirst}
\usepackage[colorlinks]{hyperref}
\usepackage[dvipsnames]{xcolor}
\usepackage{xcolor}
\usepackage{mathtools}
\DeclarePairedDelimiter{\ceil}{\lceil}{\rceil}

\newcommand{\tr}[1]{{\text{Tr}[{#1}]}}

\renewcommand\vec{\mathbf}

\definecolor{aqua}{RGB}{69,139,116}

\usepackage{subfigure}
\usepackage{algorithm}
\usepackage{algorithmicx}
\usepackage{algpseudocode}
\usepackage{amsmath}
\usepackage{verbatim}
\usepackage{makecell}
\usepackage{lettrine}
\usepackage[normalem]{ulem}

\newtheorem{theorem}{Theorem}
\newtheorem{lemma}{Lemma}

\begin{document}
\title{Quantum State Compression Shadow}

\author{Chen Ding}
\affiliation{Henan Key Laboratory of Quantum Information and Cryptography, Zhengzhou, Henan 450000, China}
\author{Xiao-Yue Xu}
\affiliation{Henan Key Laboratory of Quantum Information and Cryptography, Zhengzhou, Henan 450000, China}
\author{Shuo Zhang}
\affiliation{Henan Key Laboratory of Quantum Information and Cryptography, Zhengzhou, Henan 450000, China}
\author{Wan-Su Bao}
\email{bws@qiclab.cn}
\affiliation{Henan Key Laboratory of Quantum Information and Cryptography, Zhengzhou, Henan 450000, China}
\affiliation{Hefei National Laboratory, University of Science and Technology of China, Hefei 230088, China}
\author{He-Liang Huang}
\email{quanhhl@ustc.edu.cn}
\affiliation{Henan Key Laboratory of Quantum Information and Cryptography, Zhengzhou, Henan 450000, China}
\affiliation{Hefei National Laboratory, University of Science and Technology of China, Hefei 230088, China}
\date{\today}

\date{\today}

\begin{abstract}

\textbf{Quantum state readout serves as the cornerstone of quantum information processing, exerting profound influence on quantum communication, computation, and metrology. In this study, we introduce an innovative readout architecture called Compression Shadow (CompShadow), which transforms the conventional readout paradigm by compressing multi-qubit states into single-qubit shadows before measurement. Compared to direct measurements of the initial quantum states, CompShadow achieves comparable accuracy in amplitude and observable expectation estimation while consuming similar measurement resources. Furthermore, its implementation on near-term quantum hardware with nearest-neighbor coupling architectures is straightforward. Significantly, CompShadow brings forth novel features, including the complete suppression of correlated readout noise, fundamentally reducing the quantum hardware demands for readout. It also facilitates the exploration of multi-body system properties through single-qubit probes and opens the door to designing quantum communication protocols with exponential loss suppression. Our findings mark the emergence of a new era in quantum state readout, setting the stage for a revolutionary leap in quantum information processing capabilities.}

\end{abstract} 

\maketitle

In the realm of quantum physics, quantum measurement emerges as a pivotal and distinctive element. Its paramount role resides in the capacity to extract essential information from quantum states, enabling us to observe the fascinating phenomena of quantum physics, including measurement-induced phase transitions~\cite{ai2023measurement,koh2023measurement,choi2020quantum,skinner2019measurement}. Moreover, it serves as a prerequisite for harnessing the quantum power, with phenomena like stochastic collapse forming the security foundation for quantum key distribution~\cite{xu2020secure,bennett2014quantum}. Thus, advancements in measurement techniques are poised to revolutionize the development of quantum information processing. For instance, classical shadow~\cite{huang2020predicting,huang2022learning,zhao2021fermionic,elben2023randomized,sack2022avoiding,xu2023circuit, ding2023noise} methods have been proffered as a potent tool for extracting numerous properties from quantum many-body systems with a significantly fewer measurements. This innovation showcases the potential for more efficient utilization of quantum resources in exploring complex quantum systems. Additionally, the introduction of readout error mitigation methods~\cite{cai2023quantum,endo2018practical,temme2017error,strikis2021learning,takagi2022fundamental,huggins2021virtual,li2017efficient} seeks to surmount inaccuracies in quantum state information extraction caused by readout errors. These efforts result in a substantial enhancement of the capabilities of near-term quantum computing devices, marking a significant stride towards utility of quantum computing before fault tolerance. A notable example is the successful simulation of a 127-qubit transverse-field Ising model achieved through error mitigation techniques~\cite{kim2023evidence}. The confluence of these advancements underlines the depth of impact that refined measurement techniques can have on the multifaceted landscape of quantum information.

Here, we present a new paradigm for quantum state readout from a distinct perspective, introducing the concept of quantum state compression shadow (CompShadow). The primary goal is to compress the original quantum state into a collection of single-qubit states while ensuring the faithful recovery of the initial state's information, including amplitudes or the expectation values of specific observables, through measurement on these compressed qubits. Importantly, CompShadow introduces a plethora of new functionalities, augmenting quantum information processing capabilities: 1) It notably addresses the challenging issue of correlated measurement errors in multi-qubit systems. As quantum processors scaling up, the coupling and crosstalk between qubits become significant, leading to a globally correlated error channel with, in principle, an exponential cost for noise suppression~\cite{geller2021conditionally,nachman2020unfolding,bravyi2021mitigating,nation2021scalable}. In contrast, CompShadow simplifies large-scale quantum state measurements by measuring only a  single-qubit, thereby elegantly removing correlated errors. Numerical simulations reveal that CompShadow outperforms mainstream readout error mitigation methods such as tensor product noise inversion (TPN)~\cite{geller2020rigorous}, unfolding~\cite{nachman2020unfolding}, and model-free method~\cite{vandenberg2022modelfree} on the near-term quantum device, \textit{Zuchongzhi} 2.1~\cite{zhu2021quantum,wu_strong_2021,wu_strong_2021}. Moreover, as gate accuracy and qubit number increase, the advantage of CompShadow becomes more pronounced. 2) CompShadow facilitates revealing the properties of multi-qubit system using a single qubit prober, such as entanglement entropy and local density, potentially playing a crucial role in many-body quantum simulations. 3) It also exhibits exponential resilience against loss in quantum state transmission, significantly enhancing communication efficiency. These findings underscore CompShadow as a novel paradigm for quantum state readout, poised to markedly enhance the capabilities of contemporary quantum computing, communication, and metrology. 

\noindent\textbf{\large{Results}}

\noindent\textbf{Framework}

\noindent\textbf{\textsf{\small{Compression shadows.}}} Given a $n$-qubit quantum state $\rho$, CompShadows  is a collection of compressed quantum states $\{\sigma_1,\sigma_2,...,\sigma_{2^n-1}\}$, obtained by applying a series of compression circuits $U_j$ to $\rho$, retaining the first qubit, and discarding the remaining qubits. Mathematically,
\begin{align}
    \sigma_j=\text{Tr}_{\{2,...,n\}}(U_j\rho U_j^\dagger),\  j=1,...,2^n-1,
    \end{align}
where $U_j$ represents a hardware-efficient shallow quantum circuit that requires at most $2(n-1)$ nearest-neighbor two-qubit gates, composed as
\begin{align}\label{eqn:uj}
    U_j=\prod^{1}_{i=n-1}\text{CNOT}^{\beta_{ij}}(i+1,i)\text{CNOT}^{\alpha_{ij}}(i,i+1),
    \end{align}
where CNOT$(i,i+1)$ is an X gate on the $i+1$-th qubit controlled by the $i$-th qubit. The exponents $\alpha,\beta$ of the gates are
\begin{align}
\alpha_{ij}&=\begin{cases}
1,\quad j\geq 2^{i},\\
0,\quad \text{else},
\end{cases}\label{eqn:alpha}\\
\beta_{ij}&=\begin{cases}
1,\quad j\geq 2^{i},\ j-2^i+1 (\text{mod}\ 2^i)\leq 2^{i-1}\\
0,\quad \text{else}.
\end{cases}\label{eqn:beta}
\end{align}

\begin{figure*}[t]
\begin{center}
\includegraphics[width=1\linewidth]{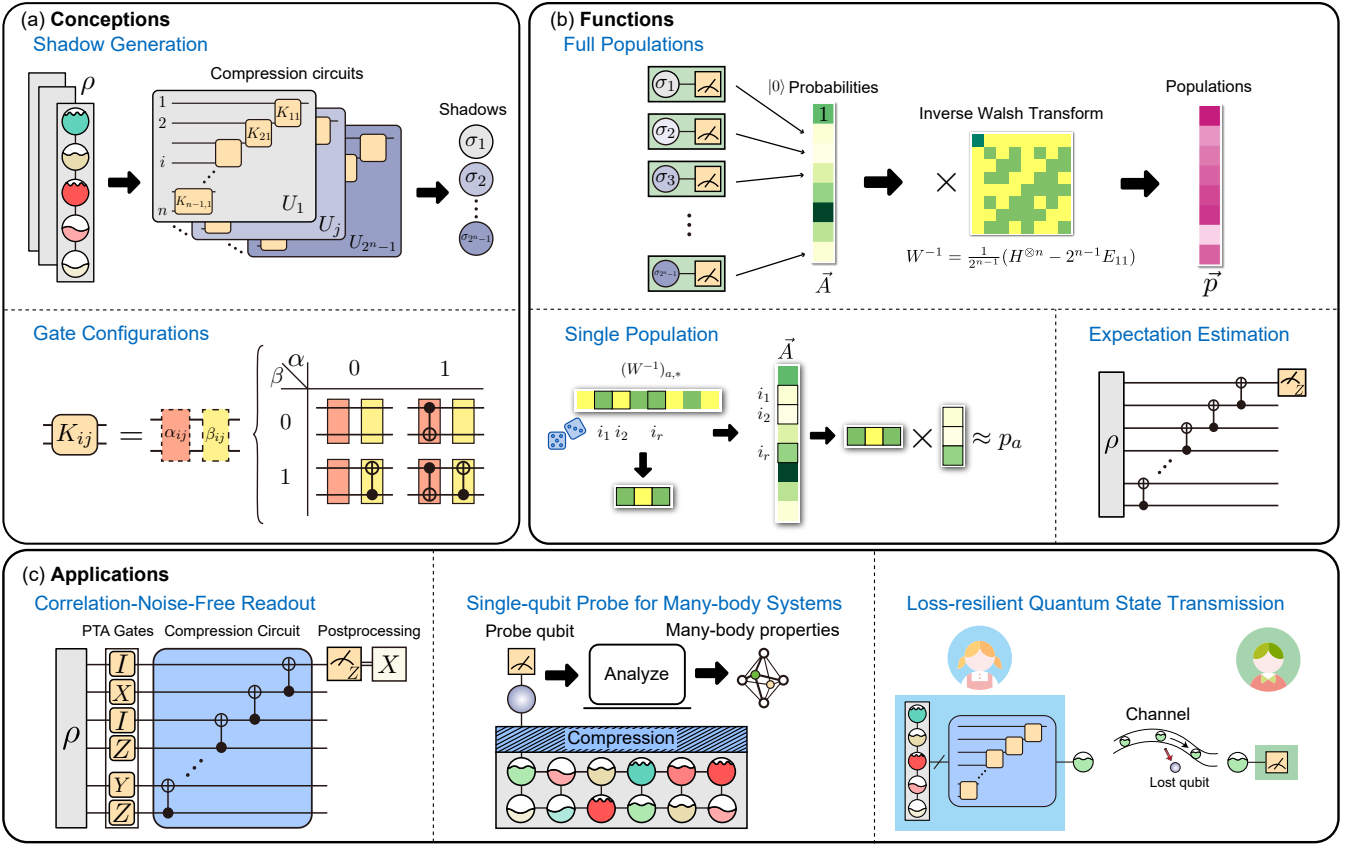}
\end{center}
\caption{\textbf{The architecture and illustrative example applications of compression shadow (CompShadow).} (a) Generation of CompShadows. Utilizing $2^n-1$ compression circuits on the $n$-qubit state $\rho$, $2^n-1$ independent CompShadows are produced. These circuits are composed of information transmission modules $K_{ij}$ between $n-1$ pairs of adjacent qubits. Each module consists solely of CNOT gates, of which the arrangements vary in each circuit $U_j$, depending on the parameters $\alpha_{ij}$ and $\beta_{ij}$ that we specifically configure in Equation~(\ref{eqn:alpha},\ref{eqn:beta}). (b) The basic functionalities of CompShadows. After measuring the $\ket{0}$ probabilities of the CompShadows, the full populations can be obtained by applying an inverse Walsh transform. Additionally, through the use of subsampling techniques~\cite{tang2019quantuminspired,andras2018quantum,tang2021quantum,ding2022quantuminspired}, it is possible to estimate individual populations with polynomial shadows. Lastly, the estimation of observable expectations can be achieved using just a single shadow. (c) The illustrative example applications of CompShadow. Leveraging the unique characteristics of requiring measurement on only a single-qubit, we can achieve high-fidelity quantum state readout, completely unaffected by correlated readout errors. Furthermore, CompShadow can serve as a single-qubit probe, allowing the exploration of properties in multi-body quantum systems. Finally, the shadows can be used for quantum state transmission, exhibiting strong resilience in the face of lossy channels.
}
\label{fig:flow}
\end{figure*}

The compression process, also shown in Fig.~\ref{fig:flow}a, achieves a special Walsh transform from the input $\rho$ to the CompShadows, as asserted in Theorem~\ref{thm:main} (See Supplementary Note~1 for the proof). This connection enables us to acquire information about $\rho$ by measuring the CompShadows, as outlined below. 
\begin{theorem}\label{thm:main} Let $A_0=1$, $A_{j}:=\text{Tr}(\sigma_j\ket{0}\bra{0})$ be the $\ket{0}$-probability of the $j$-th CompShadow $\sigma_j$, and $p_i:=\text{Tr}(\rho\ket{i}\bra{i})$ be the amplitude population of $\rho$, 
\begin{align}\label{eqn:AWp}
\vec{A}&=W\vec{p}
\end{align}
in which $W=(H^{\otimes n}+E)/2$ is the 01-valued Walsh-Hadamard Transform (WHT) matrix, Hadamard matrix $H=\begin{pmatrix}
1&1\\1&-1
\end{pmatrix}$ and $E$ is the matrix of ones.
\end{theorem}

\noindent\textbf{\textsf{\small{Amplitude population estimation.}}} Using CompShadows, the full populations of $\rho$ can be recovered, or a specific population.
According to Eqn.~(\ref{eqn:AWp}), obtaining the full amplitude vector $\vec{p}$ is straightforward, as
\begin{align}
\vec{p}&=W^{-1}\vec{A}\label{eqn:pWA}\\
&=\frac{1}{2^{n-1}}(H^{\otimes n}-2^{n-1}E_{00})\vec{A},\label{eqn:Winv}
\end{align}
in which $E_{00}=(\delta_{i0}\delta_{j0})_{ij}$ is the matrix where only the element in the first row and first column is a non-zero one. Thus, by performing only single-qubit measurements on CompShadows (see the upper half of Fig.~\ref{fig:flow}b), the population vector of $\rho$ can be recovered.

Recovering the full population vector demands all the $2^n-1$ shadows, while estimating a single population $p_a$ for arbitrary $\ket{a}$ requires only a polynomial number of shadows. Note that $p_a$ is the inner product between the $a$-th row $(W^{-1})_{a,*}$ of inverse Walsh matrix $W^{-1}$ and the shadow's $\ket{0}$-probability vector $\vec{A}$. Borrowing concepts from quantum-inspired techniques, specifically the Inner Product Estimation (IPE) algorithm~\cite{tang2019quantuminspired,andras2018quantum,tang2021quantum,ding2022quantuminspired}, allows us to efficiently estimate the inner product $p_a$ through subsampling on these vectors. For specific details, refer to the Methods section. Here, we provide a concise overview of the methodology. As illustrated in the lower-left corner of Fig.~\ref{fig:flow}b, we randomly sample indices $i_1, i_2, ..., i_r$ according to a distribution about $(W^{-1})_{a,*}$, where $r$ depends polynomially on the system size $n$ and the tolerance error of the estimation. These indices then guide us to query $(W^{-1})_{a,i_1}, (W^{-1})_{a,i_2}, ..., (W^{-1})_{a,i_r}$ through direct calculation, and $A_{a,i_1}, A_{a,i_2}, ..., A_{a,i_r}$ by measuring the corresponding shadows. The final inner product is obtained by calculating the median of means of an unbiased estimator constructed from these elements. Importantly, this approach allows us to efficiently estimate the inner product $p_a$ for arbitrary $\ket{a}$. The required number of shadows, along with the number of measurement shots, is detailed in the following theorem (See Supplementary Note~2 for the proof).

\begin{theorem}\label{thm:pop}
The CompShadow population estimation method necessitates $O(\frac{1}{\epsilon^{2}}\log(\frac{1}{\eta}))$ shadows and $O(\frac{1}{\epsilon^{4}}\log(\frac{1}{\eta}))$ measurement shots to approximate the population $p_a$ for any bitstring $\ket{a}$ within an error margin of $\epsilon$ and a success probability of $1-\eta$.
\end{theorem}

\noindent\textbf{\textsf{\small{Expectation estimation.}}} Without loss of generality, we outline the method to estimate $\braket{Z^{\otimes n}}$ for $\rho$ using CompShadow, since any observable can be decomposed into a sum of Pauli observables, and these observables can be further converted to $Z^{\otimes n}$ through single-qubit rotations. The estimation of $\braket{Z^{\otimes n}}$ can be efficiently accomplished using just one shadow and the same number of measurement shots as a direct readout, where direct readout refers to the direct measurement of $\rho$.

Based on Eqn.~(\ref{eqn:pWA}), we have
\begin{align}
\braket{Z^{\otimes n}}&=(\text{Diag}(Z^{\otimes n}),\vec{p})\label{eqn:diagZ}\\
&=(1,-1)^{\otimes n}W^{-1} \vec{A}\nonumber\\
&=(-1,0,...,0,2)\vec{A}\nonumber\\
&=2A_{2^n-1}-1,\nonumber
\end{align}
which means $\braket{Z^{\otimes n}}$ of $\rho$ can be estimated by the $\braket{Z}$ expectation of its $2^n-1$-th CompShadow, as

\begin{align}
\tr{\rho Z^{\otimes n}}&=\tr{U_{2^n-1}\rho U_{2^n-1}^\dagger Z\otimes I^{\otimes n-1}}\nonumber\\
&=\tr{\sigma_{2^n-1}Z},\label{eqn:Zrho}
\end{align}
where 
\begin{align}
    U_{2^n-1}&=\prod^{1}_{i=n-1}\text{CNOT}(i+1,i)\label{eqn:ulast}
    \end{align}
is a shallow circuit with a depth of $n-1$, as illustrated in lower-right corner of Fig.~\ref{fig:flow}b. Theorem~\ref{thm:obs} states that the quantum state copy resources required for expectation value estimation are the same for both CompShadow readout and direct readout. Refer to its proof in Supplementary Note~2.

\begin{theorem}\label{thm:obs}
The CompShadow expectation estimation method requires $O(\frac{1}{\epsilon^2})$ copies of the state to estimate the expectation of any Pauli observable within an error of $\epsilon$.
\end{theorem}

\noindent\textbf{Illustrative example applications}

\begin{figure}[!tbp]
\begin{center}
\includegraphics[width=\linewidth]{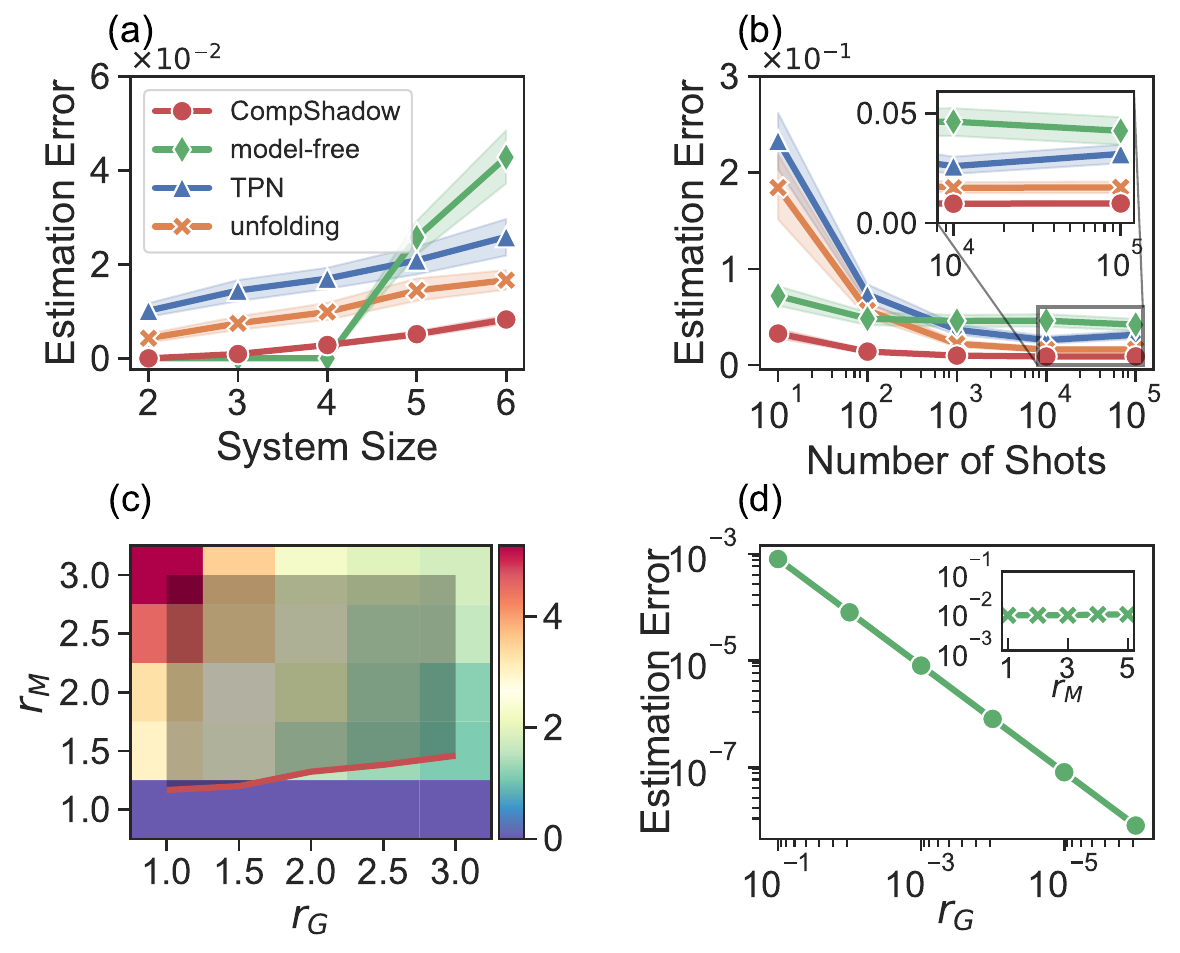}
\end{center}
\caption{\textbf{Comparative error analysis of four readout error mitigation methods.} The noise parameters are set based on typical values of \textit{Zuchongzhi} 2.1~\cite{zhu2021quantum,wu_strong_2021}. (a) Estimation errors of the four methods (CompShadow, model-free, TPN, and unfolding) with varying system size, given infinite measurement shots. The CompShadow method typically exhibits the lowest estimation errors, and its advantage becomes more pronounced with the growth of the system. (b) Estimation errors of the four methods under different measurement shots. The CompShadow method achieves the minimum estimation error with the fewest number of measurements. (c) The estimation error ratio between the unfolding method and the CompShadow method is depicted, considering amplified gate error $\widetilde{\epsilon_\text{G}}=r_G\times\epsilon_\text{G}$ and measurement error $\widetilde{\epsilon_\text{M}}=r_M\times\epsilon_\text{M}$, where 
$r_G$ ($r_M$) represents the amplification factor, and 
$\epsilon_\text{G}$ ($\epsilon_\text{M}$) accounts for gate and readout errors based on typical values from \textit{Zuchongzhi} 2.1~\cite{zhu2021quantum,wu_strong_2021,wu_strong_2021}. The red curve represents contours where both methods exhibit the same error. The shaded region highlights the superiority of CompShadow over the unfolding method.
(d) The estimation error of CompShadow versus the reduction in gate error or the amplification of measurement error, suggesting the performance of CompShadow will continue to improve as the gate noise $\epsilon_\text{G}$ decreases, and it remains steady with increasing measurement noise $\epsilon_\text{M}$.  This suggests that the performance of CompShadow will continue to improve as the gate noise $\epsilon_\text{G}$ decreases and will not vary with an increase in measurement noise $\epsilon_\text{M}$. All subplots depict the average results obtained from 100 independent repeated experiments under random computational basis states. The uncertainty bands around the curves represent the standard error of the mean.}
\label{fig:acc}
\end{figure}

\begin{figure}[t]
\begin{center}
\includegraphics[width=1\linewidth]{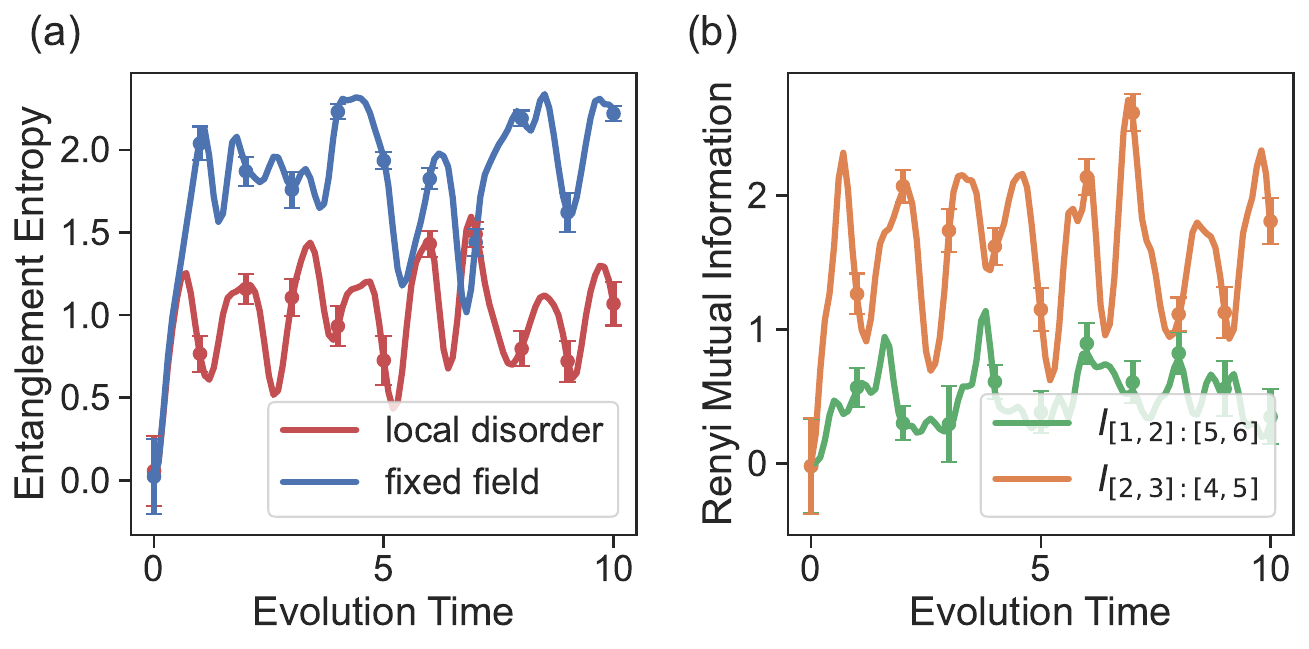}
\end{center}
\caption{\textbf{CompShadow-probed entanglement evolution of a six-qubit long-range XY model.} The system Hamiltonian is $H=\hbar\sum_{i<j}J_{ij}(\sigma^+_i\sigma^-_j+\sigma^-_i\sigma^+_j)+\hbar B\sum_j\sigma^z_j$, where $\hbar=1$ and $J_{ij}=(j-i)^{-2}$. (a) The entanglement entropies of states evolved under systems with and without random disorder. The transverse field strength $B$ is fixed at 10 for the fixed field or sampled randomly from a uniform distribution on [7,13] for the local disorder. (b) The R\'{e}nyi Mutual Information (RMI) of two pairs of subsystems during the evolution of the Hamiltonian with disorder. In these subfigures, the lines represent the theoretical values. The dots are averages of CompShadow estimation results from 100 independent experiments. The error bars on each dot represent the standard error of the mean. We find the probed value of CompShadow faithfully reflect the evolution of entanglement in the many-body system.}
\label{fig:plot_probe}
\end{figure}

\begin{figure*}[t]
\begin{center}
\includegraphics[width=0.9\linewidth]{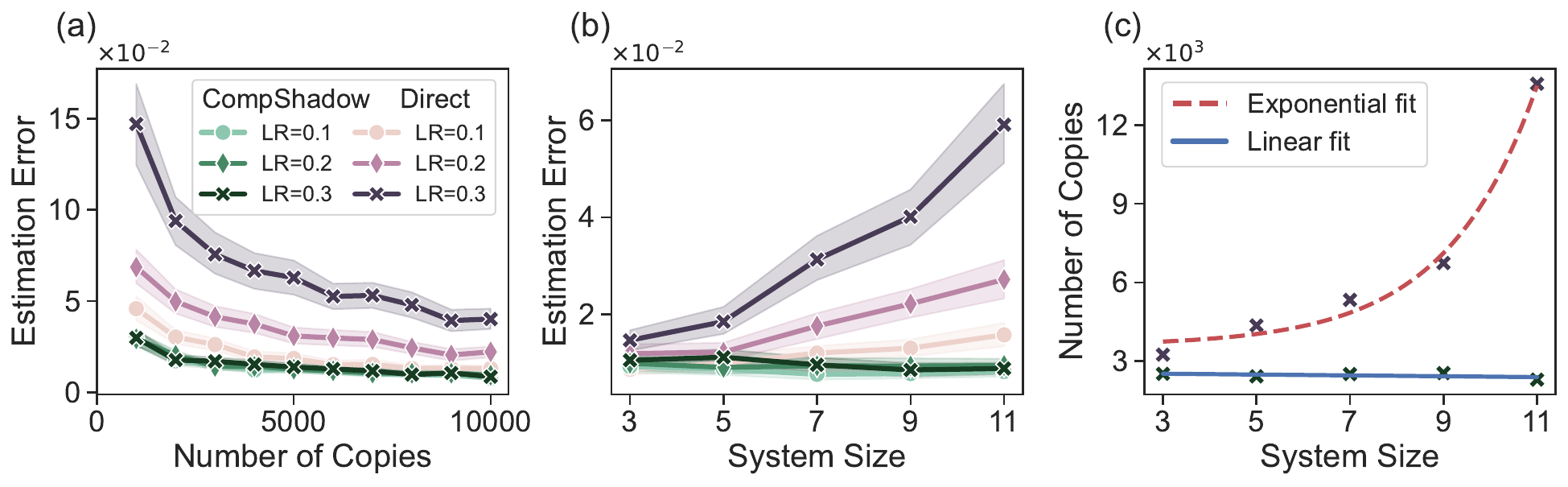}
\end{center}
\caption{\textbf{The errors of direct communication and CompShadow communication in the lossy channel.} In each experiment, copies of GHZ states are sent from Alice to Bob. Due to the presence of a lossy channel, the number of received state copies by Bob decreases, introducing additional errors into his observable estimation task. Simulations were conducted for different loss rates (LR). (a) The errors curves with varying number of copies sent by Alice, given a $9$-qubit state. (b) The errors curves with varying system size, given $10000$ copies of a state sent by Alice. (c) Required copies of GHZ states sent by Alice to achieve an estimation error less than $0.01$, varies with the system size, considering a lossy channel with a loss rate of 0.3. We employ an exponential function for the direct protocol and a linear function for the CompShadow protocol to fit the curves. In these subfigures, the markers indicate the average values from 100 independent experiments, while the bars represent the standard errors of the means. It is evident that there is a consistent exponential reduction in the required copy numbers for CompShadow communication as the system size increases, compared to direct communication.}
\label{fig:photon_loss}
\end{figure*}

\noindent\textbf{\textsf{\small{Correlation-noise-free readout.}}} Qubit readout is generally the most error-prone operation, degrading the performance of near-term quantum devices. While fully suppressing readout noise usually requires exponential calibration complexity due to the presence of correlated readout noise, CompShadow readout naturally overcomes this issue, as CompShadow readout only requires single-qubit readout. Nevertheless, CompShadow readout introduces additional quantum circuit, referred to as the compression circuit, which brings extra noise. Fortunately, we can apply randomized compiling (RC)~\cite{geller2013efficient,wallman2016noise,cai2019constructing,hashim2021randomized} to the noisy compression circuits to mitigate the gate noises  (see Methods for detail). This approach reduces the circuit errors to an approximated Pauli error channel while preserving the ideal functionality of the circuits. Consequently, the noisy observable expectation will be nearly a constant multiple of the ideal one. 

Take the estimation of $\braket{Z^{\otimes n}}$ for example. Denote $\sigma'_{2^n-1}=\text{Tr}_{\{2,...,n\}}(\mathcal{U}_{\text{RC}}(\rho))$ as the ${2^n-1}$-th shadow of $\rho$, and $\delta^{\prime}_{2^n-1}=\text{Tr}_{\{2,...,n\}}(\mathcal{U}_{\text{RC}}(\ket{0}\bra{0}^{\otimes n}))$, where RC convert the compression circuit $U_{2^n-1}$ to channel $\mathcal{U}_{\text{RC}}$. Then we estimate the error-mitigated observable expectation as
\begin{align}
\frac{\tr{\sigma'_{2^n-1}Z}}{\tr{\delta'_{2^n-1}Z}}.\label{eqn:miti}
\end{align}.

To demonstrate the advantage of CompShadow readout over conventional readout error mitigation methods, such as TPN~\cite{geller2020rigorous}, unfolding~\cite{nachman2020unfolding}, and model-free readout-error mitigation~\cite{vandenberg2022modelfree}, we compare the estimation errors $|\braket{Z^{\otimes n}}_\text{estimated}-\braket{Z^{\otimes n}}_\text{ideal}|$ of these methods on random computational basis states, where $\braket{Z^{\otimes n}}_\text{estimated}$ ($\braket{Z^{\otimes n}}_\text{ideal}$) is the estimated (ideal) observable expectation. In our simulation, we set the number of circuit instances for both CompShadow and model-free to $4n$. The single-qubit and  two-qubit gate errors, as well as the $T_1$, are based on typical values of \textit{Zuchongzhi} 2.1~\cite{zhu2021quantum,wu_strong_2021} and are set at 0.16\%, 0.6\%, and $26.5\mu\text{s}$, respectively. The readout errors arise from our calibration data (see Supplementary Note~3), which includes correlated errors in the error matrix, with an average error denoted as 2.22\%.

Figure~\ref{fig:acc} illustrates the simulation results. In Fig.~\ref{fig:acc}a, we examine the relationship between estimation errors and the system size $n$. The number of measurement shots is considered infinite to avoid the influence of statistical errors. As anticipated, the estimation errors for all four methods increase as the system size grows. Notably, as the system size expands, the superiority of CompShadow readout becomes increasingly evident among the four methods. For instance, for $n=6$, the errors of CompShadow readout are $55.3\%, 32.3\%, 18.5\%$ of those for unfolding, TPN, and model-free, respectively. The performance of the model-free method significantly degrades with the increasing system size, as the number of circuit instances we employ ($4n$) is much lower than the theoretically required quantity for the model-free method to function optimally, which is $2^n$. In Fig.~\ref{fig:acc}b, we maintain the system size at $n=6$ and vary the number of measurement shots. The errors for all methods decrease with the number of measurement shots and gradually converge. Notably, CS requires the fewest number of measurement shots to achieve low estimation error, and it converges most rapidly. Figure~\ref{fig:acc}a,b underscore the evident advantage of CS over other methods in mitigating measurement errors and its robust scalability concerning the number of qubits and measurement shots.

Next, we fix the system size at $n=6$ and investigate the performance of CompShadow readout under higher and lower noise scenarios, with infinite measurement shots. In Fig.~\ref{fig:acc}c, we systematically amplify gate error $\epsilon_G$ and readout error $\epsilon_M$. The contour plot illustrates the ratio of estimation errors between the unfolding and CompShadow methods (two methods for the victory in Fig.~\ref{fig:acc}a,b.), with contours of ratio $1$ highlighted in red. We observe a substantial advantage of CompShadow readout in scenarios with high measurement error and low gate error. Remarkably, even as gate error $\epsilon_\text{G}$ is magnified to three times large, as long as there is a slight increase in measurement error $\epsilon_\text{M}$ (to about $50\%$ larger), the advantage of CompShadow readout persists. Therefore, CompShadow readout can serve as a valuable complement for readout-error mitigation methods, particularly in systems characterized by high readout errors and low gate errors. In Fig.~\ref{fig:acc}d, we fix measurement error $\epsilon_\text{M}$, revealing that the estimation error of CompShadow readout decreases polynomially with gate error $\epsilon_\text{G}$. When the gate error is reduced to $10^{-6}$ of its original level, the estimation error of CompShadow can be minimized to about $8\times 10^{-9}$. Conversely, when fixing gate error $\epsilon_\text{G}$ and increasing measurement error $\epsilon_\text{M}$, the estimation errors of CompShadow for all cases are approximately 0.008, with some minor fluctuations mainly attributed to the randomness in the RC circuit.

\noindent\textbf{\textsf{\small{Single-qubit probe for many-body systems.}}} The application of CompShadow readout for probing the properties of many-body systems not only effectively mitigates readout errors, as discussed earlier, but also introduces a distinctive feature. The compressed individual qubit, referred to as a single-qubit probe, can either be an integral part of the system or an auxiliary qubit derived from another system (see the Supplementary Note~5). This feature allows for a scenario where a particular many-body system is physically challenging to measure, we can leverage an easily measurable system, implementing the CompShadow compression circuit, to interact with it. By reading the easily measurable system, we can gain insights into the properties of the target system. In this context, we present several examples, such as utilizing CS to measure second-order R\'{e}nyi entropy and R\'{e}nyi mutual information (RMI), defined as
\begin{align}
S^{(2)}(\rho)&=-\log_2\text{Tr}(\rho^2),\\
I^{(2)}(\rho_A:\rho_B)&=S^{(2)}(\rho_A)+S^{(2)}(\rho_B)-S^{(2)}(\rho_{AB}),
\end{align}
and local density~\cite{cramer2008exploring,zhu2022observation} (refer to the Supplementary Note~4 for specific details).

Combining CompShadow readout with the randomized measurement (RM) methods~\cite{brydges2019probing,elben2023randomized} allows for the single-qubit probing of the second-order R\'{e}nyi entropy. The main idea is that given copies of the state $\rho$, for $M$ rounds, we apply a layer of random single-qubit Clifford gates on the system $\rho$. The outcome states, denoted as $U^{(i)}\rho U^{(i)\dagger}, i=1,...,M$, are then compressed into shadows and measured. For a more detailed process, see the Methods.

Numerical simulations are conducted on a 6-qubit long-range XY model~\cite{porras2004effective} defined by the Hamiltonian $H=\hbar\sum_{i<j}J_{ij}(\sigma^+_i\sigma^-_j+\sigma^-_i\sigma^+_j)+\hbar B\sum_j\sigma^z_j$, where $\hbar=1$ and $J_{ij}=(j-i)^{-2}$. The transverse field is fixed at $B=10$, and local disorder potentials $B_j$ are uniformly sampled from the interval $[7,13]$. In Fig.~\ref{fig:plot_probe}a, the evolution of entanglement entropies with and without random disorder is presented. The entanglement entropy of both systems initially undergoes a rapid increase, followed by sustained oscillations. The system without disorder generally exhibits a higher degree of entanglement. Fig.~\ref{fig:plot_probe}b illustrates the RMI for two pairs of subsystems during the evolution of the perturbed Hamiltonian. The distant subsystems display a lower degree of RMI compared to the adjacent ones. In both cases, the CompShadow method (dots) generally align with the theoretical predictions (lines). These results showcase the capability of CompShadow in measuring the entanglement structure of many-body systems.

\noindent\textbf{\textsf{\small{Loss-resilient quantum state transmission.}}} Due to CompShadow compressing quantum states into a single qubit, transmitting the state after CompShadow compression through a lossy channel is more robust against errors compared to transmitting the original state. This characteristic bears significant implications for quantum communication or applications in a quantum internet.

Here, we present a proof-of-concept example. Let's consider Alice transmitting an $n$-qubit GHZ state to Bob, who aims to estimate the expectation value of $Z^{\otimes n}$. Alice has two options: she can directly send the $n$ qubits (referred to as "direct communication"), or she can compress the quantum state into a single-qubit shadow using the compression circuit $U_{2^n-1}$ in Eqn.~(\ref{eqn:ulast}) (referred to as "CompShadow communication"). In both scenarios, the loss of a single qubit is considered a transmission failure. In the transmission channel, each qubit faces an independent and fixed loss rate.

In Fig.~\ref{fig:photon_loss}a,b, we fix the system size and the number of copies of the quantum state transmitted by Alice, and then vary the number of copies and system size, observing the estimation error of CompShadow communication and direct communication. Notably, due to the compression into a single qubit, CompShadow demonstrates a marked advantage against lossy errors. The estimation error in the CompShadow protocol is significantly lower, and this advantage becomes more pronounced with decreasing copy numbers and increasing system sizes. The primary reason for this lies in the CompShadow protocol, where Bob obtains copies ($(1-r)n_{c}$) with an exponential growth rate compared to direct communication ($(1-r)^{n_{q}}n_{c}$), where $r$ is the loss rate, $n_q$ and $n_c$ is the system size and number of copies Alice sent. Finally, in Fig.~\ref{fig:photon_loss}c, we set the target to achieve an estimation error less than $0.01$ and compare the minimum number of copies required for both protocols under different system sizes. As anticipated, the CompShadow protocol requires exponential fewer copies compared to the direct protocol with the system size increasing. When the system size reaches $n=11$, direct communication requires approximately six times more number of copies than CompShadow communication. This series of experiments clearly illustrates the advantages of the CompShadow protocol in transmitting quantum states through lossy channels, not only in reducing copy numbers but also in improving measurement accuracy.

\noindent\textbf{\large{Discussion}}

The CampShadow efficiently compresses a large-scale quantum state into individual qubits, introducing only some additional hardware-efficient shallow quantum circuits with at most $2(n-1)$ nearest-neighbor two-
qubit gates for $n$ qubits. This approach is highly practical for near-term quantum devices. The compression process can be conceptualized as a lossless quantum auto-encoder, as we can recover information such as amplitudes and observable expectations of the original quantum state through measurements on the compressed single qubit. The characteristics of the CampShadow provides a robust impetus for the advancement of quantum technologies, such as improved readout accuracy and an exponential suppression of losses in quantum state transmission. Besides, we would like to share some additional insights:

Firstly, due to the fact that CampShadow readout only requires measurements on individual qubits, errors in single-qubit measurements can be nearly completely eliminated using the TPN error mitigation method. This may imply a new direction in hardware development, where the focus can shift towards enhancing the fidelity of quantum gates without the need to simultaneously optimize multi-qubit readout performance. This shift could potentially significantly reduce the complexity of designing and manufacturing high-performance hardware. 

Secondly, CompShadow is a versatile framework with strong flexibility. In Supplementary Note~5, we demonstrate that the compressed qubit can exist either within the system or as an external ancillary qubit. Additionally, the number of qubits after compression can be adjusted to reduce the depth of the compression circuit.

Thirdly, the application of the single-qubit prober can be considered a form of weak measurement. The unmeasured portion of the system still contains information about the original system, potentially exploitable to unveil more quantum properties.

Finally, the CompShadow readout involves a Walsh transform from quantum states to classical vectors. Considering the wide-ranging applications of Walsh transforms in fields like image compression, signal processing, and cryptography, the compression circuit of our method has the potential to catalyze the quantization of related applications.

\noindent\textbf{\large{Methods}}

\noindent\textbf{Amplitude population estimation}

We present the method for estimating an amplitude population $p_a$, utilizing only a polynomial number of shadows. Note that $p_a=((W^{-1})_{a,*},\vec{A})$ is a specific inner product between vectors of exponential length. The avoidance of exponential calculation complexity is accomplished through the inner product estimation method (IPE)~\cite{tang2019quantuminspired,andras2018quantum,tang2021quantum,ding2022quantuminspired}. Generally, to estimate the inner product between vector $A$ and $B$, the method repeatedly samples the length-square distribution $\mathbb{P}(i)=\frac{|A_i|^2}{\|A\|^2}$ of $A$, generating indices $i_1,i_2,...,i_r$. Subsequently, a median of means for the values $\frac{\|A\|^2}{A_i}B_i, i=i_1,i_2,...,i_r$ is calculated as the estimation of $(A,B)$. The pseudo code in Alg.~\ref{alg:inner} outlines the parameter choice for IPE.

\begin{algorithm}[H]
\centering
\begin{algorithmic}[1]
\item[\textbf{Input:}] A sampling oracle and a query oracle for $A\in \mathbb{C}^{N}$. A query oracle for $B\in \mathbb{C}^{N}$. The Frobenius norm squared $\|A\|_F^2$. Relative error bound $\xi$ and success probability bound $1-\eta$.
\item[\textbf{Goal:}] Estimate $(A,B)$.
\State Sample $A$ for $s=\ceil{\frac{9}{\xi^2}}$ times, yielding $i_1,..., i_{s}$. Calculate the mean of $\frac{\|A\|^2}{A_{i}}B_{i}$ among these samples.
\State Repeat Step 1 for $\ceil{6\log_2(\frac{2}{\eta})}$ times and calculate the median of the $s$ outputs, denoted as $x$.
\item[\textbf{Output:}] $x$.
\end{algorithmic}
\caption{Inner Product Estimation.}
\label{alg:inner}
\end{algorithm}

The following Lemma~\ref{lemma:inner} states the complexity of the IPE algorithm. See its proof in Ref.~\cite{andras2018quantum,tang2019quantuminspired}.

\begin{lemma}\label{lemma:inner}
Suppose that we have sampling access to $A\in\mathbb{C}^{N}$ in complexity $L(A)$ and query access to $B\in\mathbb{C}^{N}$ in complexity $Q(B)$. Then we can estimate $(A,B)$ to precision $\xi\|A\|\|B\|$ with probability at least $1-\eta$ in time
\begin{align}
    O\left(\frac{\log(1/\eta)}{\xi^2}(L(A)+Q(B))\right).
\end{align}
\end{lemma}

In our case, vector $A$ corresponds to the $a$-th row of the inverse Walsh matrix $W^{-1}$, and vector $B$ represents the $\ket{0}$-probabilities $\vec{A}$ of the shadows. Referring to the input in Alg.~\ref{alg:inner}, it is essential to clarify how to sample the length-square distribution of $(W^{-1})_{a,*}$, query the elements of $\vec{A}$, and define the norm squared $\|(W^{-1})_{a,*}\|^2$. To begin, the element$A_j$ are queried by measuring the $j$-th shadow. Subsequently, leveraging on our knowledge about Walsh matrix, we can easily find
\begin{align}
    (W^{-1})^2_{ij}&=\begin{cases}
    (2^{1-n}-1)^2, &i=j=0,\\
    2^{2-2n}, &\text{else},
    \end{cases}\\
    \|(W^{-1})_{i,*}\|^2&=\begin{cases}
    1, &i=0,\\
    2^{2-n}, &\text{else}.
    \end{cases}
    \end{align}

Finally, we formulate an efficient length-square sampling algorithm for $(W^{-1})_{a,*}$ with arbitrary value of $a$, shown in Alg.~\ref{alg:row_sample}. To achieve amplitude population estimation within a precision of $\epsilon$ and a success rate of $1-\eta$, the IPE parameters are taken as $\xi=\epsilon/2$ and $\eta$. According to Lemma~\ref{lemma:inner}, the time complexity for executing IPE is $O(\frac{1}{\epsilon^{4}}\log(\frac{1}{\eta}))$. A detailed complexity analysis of the amplitude population estimation algorithm is presented in the proof of Theorem~\ref{thm:pop} in Supplementary Note~2.

\begin{algorithm}[H]
    \centering
    \begin{algorithmic}[1]
    \item[\textbf{Input:}] System size $n$. Row index $i$.
    \item[\textbf{Goal:}] Sampling an index $j$ that follows the distribution $\mathbb{P}(j)=\frac{(W^{-1})^2_{ij}}{\|(W^{-1})_{i,*}\|^2}$.
    \State If $i\neq 0$, sample $j$ from a uniform distribution in the set $\{0,1,...,2^n-1\}$. Output $j$ and end the algorithm.
    \State If $i=0$, generate a random real number $x$ from uniform distribution $[0,1]$. If $x\leq (2^{1-n}-1)^2$, return $j=0$. Otherwise, sample $j$ from a uniform distribution in the set $\{1,...,2^n-1\}$.
    \item[\textbf{Output:}] $j$.
    \end{algorithmic}
    \caption{Sampling a certain row of inverse Walsh matrix.}
    \label{alg:row_sample}
    \end{algorithm}

\noindent\textbf{Simplifying errors in the compression circuit}

To simplify the errors, including circuit errors and the single-qubit measurement errors, into a single Pauli error channel, we employ randomized compiling (RC) across the noisy compression circuit. As depicted in Fig.~\ref{fig:flow}c, before the compression circuit $U$, we randomly select a layer of Pauli gates $\otimes_{i=1}^n P^{(i)}$ to apply to the target state. After the measurements, we classically conduct $U^\dagger\otimes_{i=1}^n P^{(i)}U$, which, due to the Cliffordness of $U$, is also a layer of Pauli gates $\otimes_{i=1}^n Q^{(i)}$ to the measurement outcomes. As we only measure the first qubit, the Paulis on other qubits can be omitted. The remaining operation becomes: If $Q^{(0)}$ equals $X$ or $Y$, we flip the measured bits; otherwise, we leave it unchanged. This process is repeated in multiple rounds, the average of the obtained measurement results is taken.

Next we analyze the impact of RC on CompShadow readout. Without loss of generality, we focus on the estimation of observable expectation $Z^{\otimes n}$. Let $\tilde{\mathcal{U}}$ denote the channel of the noisy compression circuit, $\mathcal{E}_m$ represent the measurement error channel on the first qubit, $\mathcal{I}$ stand for the identity channel on each qubit. Then, the noisy expectation value of $Z^{\otimes n}$ is
\begin{align}
    \braket{Z^{\otimes n}}_{\text{noisy}}=\tr{\mathcal{E}_m\otimes \mathcal{I}^{\otimes n-1}\circ \tilde{\mathcal{U}}(\rho) Z\otimes I^{\otimes n-1}}\nonumber,
    \end{align}

Let $\mathcal{E}_c(\cdot)=\tilde{\mathcal{U}}(U^\dagger \cdot U)$, the channel $
\mathcal{E}=\mathcal{E}_m\otimes \mathcal{I}^{\otimes n-1}\circ \mathcal{E}_c$ is then the overall noise channel. As described above, in the $j$-th round, we apply the Pauli gates $\otimes_{i=1}^n P^{(i)}$ on $\rho$, and apply $U^\dagger\otimes_{i=1}^n P^{(i)}U$ on the measurement outcomes. If we iterate through all $n$-qubit Pauli operators with $4^n$ rounds, the error channel $\mathcal{E}$ is converted to a Pauli noise channel $\mathcal{E}_p(\cdot)=\sum_i\lambda_iP_i\cdot P_i$, in which $P_i$ are $n$-qubit Pauli operators~\cite{joseph2007symmetrized,hashim2021randomized}. 
In this case,
\begin{align}
\braket{Z^{\otimes n}}_{\text{twirl}}&=\tr{\mathcal{E}_p(U\rho U^\dagger) Z\otimes I^{\otimes n-1}}\nonumber\\
&=\sum_i\lambda_i\tr{P_iU\rho U^\dagger P_i Z\otimes I^{\otimes n-1}}\nonumber\\
&=\sum_i\lambda_i\tr{U\rho U^\dagger P_i^{(1)} ZP_i^{(1)}\otimes I^{\otimes n-1}}\nonumber\\
&=\sum_i\lambda_ih_i\tr{U\rho U^\dagger Z\otimes I^{\otimes n-1}}\nonumber\\
&=\sum_i\lambda_ih_i\braket{Z^{\otimes n}}_{\text{ideal}},\label{eqn:twirling_ratio}
\end{align}
in which $P_i^{(1)}$ is the Pauli operator on the first qubit of $P_i$, and
\begin{align*}
h_i=\begin{cases}
1, &P_i^{(1)}=I,Z,\\
-1, &P_i^{(1)}=X,Y.
\end{cases}
\end{align*}

Eqn.~(\ref{eqn:twirling_ratio}) suggests that the noisy expectation is a fixed multiple of the ideal one if all $4^n$ Pauli operators are iterated. The multiplier $\sum_i\lambda_ih_i$ can be estimated by measuring the $Z^{\otimes n}$ expectation for state $\ket{0}\bra{0}^{\otimes n}$ through the compression circuit, leading to the error mitigation formula as Eqn.~(\ref{eqn:miti}). In our experiments, we typically employ $4n$ random selected Pauli operators, which approximately achieve this goal. The choice of the linear number $4n$ is efficientand empirically effective, as demonstrated in Supplementary Note~3.

\noindent\textbf{Entanglement entropy probing}

We first derive a calculation formula suitable for CompShadow. Drawing from the randomized measurement (RM) methods~\cite{brydges2019probing,elben2023randomized}, the second-order R\'{e}nyi entropy of a quantum state $\rho$ can be estimated by 
\begin{equation}\label{eqn:RM}
S^{(2)}(\rho)=-\log_2\frac{2^n}{M}\sum_i\sum_{s,s'}(-2)^{-h(s,s')}\mathbb{P}(s,i)\mathbb{P}(s',i),
\end{equation}
in which $s,s'$ are summed over all measured bitstrings, $h(s,s')$ is the Hamming distance between $s$ and $s'$, and $\mathbb{P}(s,i)$ is the $s$-th amplitude population of state $U^{(i)}\rho U^{(i)\dagger}$.

Eqn.~(\ref{eqn:RM}) can be rewritten as
\begin{equation}
S^{(2)}(\rho)=-\log_2\frac{1}{M}\sum_i \vec{p}^{(i)T}D\vec{p}^{(i)},
\end{equation}
in which $\vec{p}^{(i)}$ is the amplitude population vector of $U^{(i)}\rho U^{(i)\dagger}$ in $i$-th round, and matrix $D_{ij}=2^n(-2)^{-h(i,j)}$. By Eqn.~(\ref{eqn:pWA}), we have
\begin{equation}\label{eqn:WDW}
S^{(2)}(\rho)=-\log_2\frac{1}{M}\sum_i \vec{A}^{(i)T}W^{-T}DW^{-1}\vec{A}^{(i)},
\end{equation}
in which
\begin{equation*}
W^{-T}DW^{-1}=\begin{pmatrix}
2^n&-\frac{d_1}{2}&-\frac{d_2}{2}&\cdots&-\frac{d_{2^n-1}}{2}\\
-\frac{d_1}{2}&d_1&0&\cdots&0\\
-\frac{d_2}{2}&0&d_2&\cdots&0\\
\vdots&\vdots&\vdots&\ddots&\vdots\\
-\frac{d_{2^n-1}}{2}&0&0&\hdots&d_{2^n-1}
\end{pmatrix},
\end{equation*}
and
\begin{align}\label{eqn:dj}
    d_j=3^{h(j,0)}2^{2-n}.
\end{align}

Therefore,
\begin{equation}\label{eqn:cs_ent}
S^{(2)}(\rho)=-\log_2\frac{1}{M}\sum_i\left(2^n+\sum_{j=1}^{2^n-1}d_j(A^{(i)2}_j-A^{(i)}_j)\right),
\end{equation}

Then, we utilize the IPE algorithm (Alg.~\ref{alg:inner}) to efficiently determine the sum $\sum_{j=1}^{2^n-1}d_j(A^{(i)2}_j-A^{(i)}_j)$. For the input of IPE, the sampling oracle for $\vec{d}$ is provided in Alg.~\ref{alg:sample_d}. The Frobenius norm can be found as $\|\vec{d}\|^2_F=2^{4-2n}(10^n-1)$. To query the element $A^{(i)2}_j-A^{(i)}_j$, we note the measured value of $A^{(i)}_j$ can not be directly substituted into $A^{(i)2}_j-A^{(i)}_j$, because $\mathbb{E}\tilde{A}^2\neq (\mathbb{E}\tilde{A})^2$. An unbiased estimator can be designed as
\begin{equation}
    \widetilde{A^{(i)2}_j-A^{(i)}_j}=\frac{1}{N(N-1)}\sum_{k\neq l} (1-X_k)X_l,
    \end{equation}
in which $X_1,...,X_N$ are the measurement outcomes from the shadow $\sigma_j^{(i)}$.

\begin{algorithm}[H]
    \centering
    \begin{algorithmic}[1]
    \item[\textbf{Input:}] System size $n$.
    \item[\textbf{Goal:}] Sampling an index $j$ that follows the distribution $\mathbb{P}({j})=\frac{d_{j}^2}{\|\vec{d}\|^2}=\frac{3^{2h(j,0)}}{10^n-1}$.
    \State Generate $n$ random real number $x_1,...,x_n$ from uniform distribution $[0,1]$.
    \State For $k=1,...,n$, if $x_k<0.1$, $j_k=0$, else $j_k=1$. Yield a $n$-bit binary number $j=(j_nj_{n-1}...j_1)_2$ accordingly.
    \State If $j=(0...0)_2$, go to Step 1. Otherwise, return $j$.
    \item[\textbf{Output:}] $j$.
    \end{algorithmic}
    \caption{Efficient sampling of vector $\vec{d}$.}
    \label{alg:sample_d}
    \end{algorithm}

\noindent\textbf{\large{Competing Interests}}

The authors declare no competing interests.

\noindent\textbf{\large{Author Contributions}}
H.-L. H. supervised the whole project. C. D. and H.-L. H. conceived the idea and co-wrote the paper. C. D. provided the theoretical analysis and proofs. C. D., X.-Y. X. and H.-L. H. carried out the numerical simulations and analyzed the results. All authors contributed to discussions of the results.

\noindent\textbf{\large{Acknowledgments}}

H.-L. H. is supported by the Youth Talent Lifting Project (Grant No. 2020-JCJQ-QT-030), National Natural Science Foundation of China (Grants No. 12274464).

\bibliographystyle{naturemag}
\bibliography{b}

\end{document}


\title{Supplementary Information for ``Quantum State Compression Shadow''}
\maketitle

\section{Proof for the Correctness of CompShadow}\label{sm:sec:thm1-proofs}

We provide the proof for Theorem~1 in the maintext, which asserts the relation between $\ket{0}$-probabilities of shadows and the amplitude populations of the target state, as $\vec{A}=W\vec{p}$. This is the basic theorem that claims the functionalities of CompShadow, supporting the development of the following readout algorithms. We repeat the theorem here for clarity.


\begin{theorem} Let $A_0=1$, $A_{j}:=\text{Tr}(\sigma_j\ket{0}\bra{0})$ be the $\ket{0}$-probability of the $j$-th CompShadow $\sigma_j$, and $p_i:=\text{Tr}(\rho\ket{i}\bra{i})$ be the amplitude population of $\rho$, 
    \begin{align}\label{eqn:AWp}
    \vec{A}&=W\vec{p}
    \end{align}
    in which $W=(H^{\otimes n}+E)/2$ is the 01-valued Walsh-Hadamard Transform (WHT) matrix, Hadamard matrix $H=\begin{pmatrix}
    1&1\\1&-1
    \end{pmatrix}$ and $E$ is the matrix of ones.
    \end{theorem}

\begin{figure*}[t]
\begin{center}
\includegraphics[width=\linewidth]{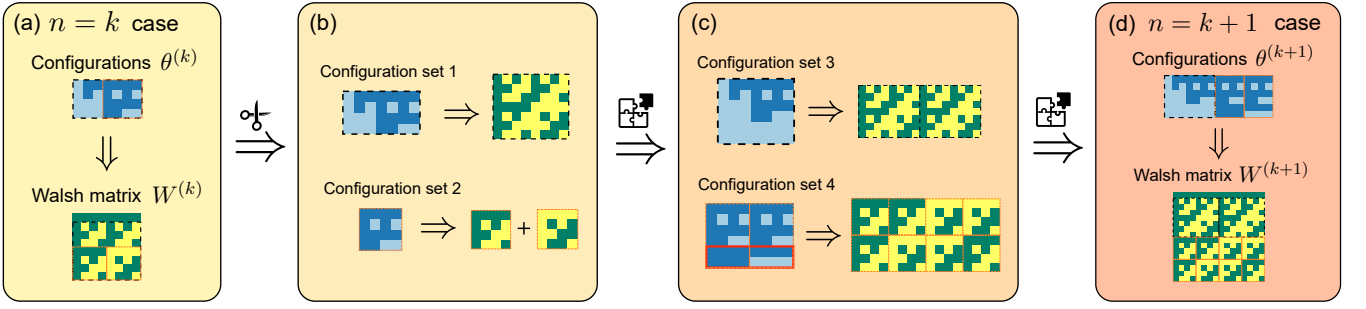}
\end{center}
\caption{\textbf{Illustration of the proof strategy of Theorem 1.} The proof of $\vec{A}=W\vec{p}$ boils down to showing the correspondence between configuration set of exponent values $\theta$ in the compression circuits and the Walsh matrix. We use the mathematical induction method to show the proposition holds for arbitrary system size $n$. (a) Suppose when the system size $n=k$, we already know the $\theta$ defined by Eqn.~(\ref{eqn:alpha}) and Eqn.~(\ref{eqn:beta}) corresponds to Walsh matrix $W^{(k)}$. (b) According the assumption for $n=k$, we yield two correspondence relation between configuration sets and submatrices of Walsh matrix $W^{(k)}$. (c) Subsequently, we can splice the correspondence relation to yield submatrices of the larger Walsh matrix $W^{(k+1)}$. The extended configuration set can be directly applied in the $k+1$-qubit compression circuits. (d) Finally, we complete the splicing and yield the proposition for system size $n=k+1$.}
\label{fig:thmalg1}
\end{figure*}

\begin{proof}[Proof of Theorem~1]
The first row of $W$ is $W_{0,*}=(1,...,1)$, which satisfies $(W_{0,*},\vec{p})=A_0=1$. By definition, for $j=1,...,2^n-1$,
\begin{align*}
A_{j}&=\text{Tr}(\sigma_j\ket{0}\bra{0})\\
&=\text{Tr}(\text{Tr}_{\{2,...,n\}}(U_j\rho U_j^\dagger)\ket{0}\bra{0})\\
&=\text{Tr}(U_j\rho U_j^\dagger\ket{0}\bra{0}\otimes I^{\otimes n-1}),
\end{align*}
in which $\ket{0}\bra{0}\otimes I^{\otimes n-1}=\text{Diag}(1,1,...,1,0,...,0)$ is diagonal. Denote $D_i=(\ket{0}\bra{0}\otimes I^{\otimes n-1})_{ii}$. Then
\begin{align*}
A_j&=\sum_{i=0}^{2^n-1} D_i \text{Tr}(U_j\rho U_j^\dagger\ket{i}\bra{i})\\
&=\sum_{i=0}^{2^n-1} D_i \text{Tr}(\rho U_j^\dagger\ket{i}\bra{i}U_j).\\
\end{align*}

Composed by CNOT gates, the unitary $U_j$ is a permutation among the computational basis states. Therefore, $U^\dagger_j\ket{i}=\ket{\pi_j(i)}$ is a basis state. Then
\begin{align}
A_j&=\sum_{i=0}^{2^n-1} D_i \text{Tr}(\rho \ket{\pi_j(i)}\bra{\pi_j(i)})\nonumber\\
&=\sum_{i=0}^{2^n-1} D_i p_{\pi_j(i)}\nonumber\\
&=\sum_{l=0}^{2^n-1} D_{\pi_j^{-1}(l)}p_l.\label{eqn:Dmat}
\end{align}

Therefore, our remaining task is to prove $D_{\pi_j^{-1}(l)}=W_{j,l}$. Denote $D_{j,l}=D_{\pi_j^{-1}(l)}$. By Eqn.~(\ref{eqn:Dmat}), $D_{j,l}\in\{0,1\}$ is the value of $\ket{0}$ probability of the $j$-th shadow with input state $\ket{l}$, which totally depends on the structure of the compression circuits. By definition in maintext, we know
\begin{align}\label{eqn:uj}
    U_j=\prod^{1}_{i=n-1}\text{CNOT}^{\beta_{ij}}(i+1,i)\text{CNOT}^{\alpha_{ij}}(i,i+1),
    \end{align}
where CNOT$(i,i+1)$ is an X gate on the $i+1$-th qubit controlled by the $i$-th qubit. The exponents $\alpha,\beta$ of the gates are
\begin{align}
\alpha_{ij}&=\begin{cases}
1,\quad j\geq 2^{i},\\
0,\quad \text{else},
\end{cases}\label{eqn:alpha}\\
\beta_{ij}&=\begin{cases}
1,\quad j\geq 2^{i},\ j-2^i+1 (\text{mod}\ 2^i)\leq 2^{i-1}\\
0,\quad \text{else}.
\end{cases}\label{eqn:beta}
\end{align}


For simplicity we let $\theta_{2i-1,j}=\beta_{ij}, \theta_{2i,j}=\alpha_{ij}$. The values of $\theta$ form a $2^n-1$-size set of $2n-2$-length configuration vectors. We apply the mathematical induction method to show the correspondence between the configuration set and the Walsh matrix, which directly implies $D=W$. The $n=2,3$ case can be validated. Supposing when $n=k$, the statement holds true, we yield the following two facts:

1) Configuration set 1 $\vec{\theta}^{(k)}_{1\sim 2k-2,1\sim 2^k-1}$ introduces the $2\sim 2^k$-th row of $W^{(k)}$, as suggested in the black frames in Fig.~\ref{fig:thmalg1}. 

2) Configuration set 2 $\vec{\theta}^{(k)}_{1\sim 2k-2,2^{k-1}\sim 2^k-1}$ introduces the $2^{k-1}+1\sim 2^k$-th row of $W^{(k)}$, as suggested in the orange frames in Fig.~\ref{fig:thmalg1}. The submatrix $W^{(k)}_{2^{k-1}+1\sim 2^k,*}$ can be further divided into two blocks: $a$. $W^{(k)}_{2^{k-1}+1\sim 2^k,1\sim 2^{k-1}}$, $b$. $W^{(k)}_{2^{k-1}+1\sim 2^k,2^{k-1}+1\sim 2^k}$, depending on the state of the last qubit. For any input basis state $\ket{l}$, if the last qubit is in state $\ket{0}$, configuration set 2 introduces block $a$. And the set introduces block $b$ if the state is $\ket{...1}$.

Then we consider the $n=k+1$ case. From the matrix of $\theta^{(k+1)}$, we can easily find the patterns of configuration set 1 and 2 as $\vec{\theta}^{(k+1)}_{1\sim 2k-2,1\sim 2^k-1}$, $\vec{\theta}^{(k+1)}_{1\sim 2k-2,2^{k}\sim 2^k+2^{k-1}-1}$, and $\vec{\theta}^{(k+1)}_{1\sim 2k-2,2^k+2^{k-1}\sim 2^{k+1}}$, as also illustrated in Fig.~\ref{fig:thmalg1}. 

For the configurations $\vec{\theta}^{(k+1)}_{*,1\sim 2^k-1}$ that includes the pattern of configuration set 1, the remaining two parameters $\theta^{(k+1)}_{2k-1,*}$ and $\theta^{(k+1)}_{2k,*}$ are all taken as zero, which suggests the $k+1$-th qubit does not interfere with others during the compression. Thus, for any basis state $\ket{l}$ in the $k+1$-qubit Hilbert space, the introduced values of $D^{(k+1)}$ form the same pattern as the $2\sim 2^k$-th row of $W^{(k)}$, whether the $k+1$-th qubit is in $\ket{0}$ or $\ket{1}$.          

For the configurations $\vec{\theta}^{(k+1)}_{*,2^{k}\sim 2^k+2^{k-1}-1}$ that includes the pattern of configuration set 2, the remaining two parameters $\theta^{(k+1)}_{2k-1,*}$ and $\theta^{(k+1)}_{2k,*}$ are all taken as one, which suggests two CNOT gates between the $k$-th and $k+1$-th qubit. For any basis state $\ket{l}$ in the $k+1$-qubit Hilbert space, the two CNOT gates transform it by
\[\ket{...00}\rightarrow\ket{...00}, \ket{...01}\rightarrow\ket{...11}, \ket{...10}\rightarrow\ket{...01}, \ket{...11}\rightarrow\ket{...10}.\]
The introduced values of $D^{(k+1)}$ thus form the pattern $a,a,b,b$.

For the configurations $\vec{\theta}^{(k+1)}_{*,2^k+2^{k-1}\sim 2^{k+1}}$ that includes the pattern of configuration set 2, the remaining two parameters $\theta^{(k+1)}_{2k-1,*}$ and $\theta^{(k+1)}_{2k,*}$ are $1,0$, which suggests CNOT$(2k,2k-1)$ between the $k$-th and $k+1$-th qubit. For any basis state $\ket{l}$ in the $k+1$-qubit Hilbert space, the CNOT gate transforms it by
\[\ket{...00}\rightarrow\ket{...00}, \ket{...01}\rightarrow\ket{...11}, \ket{...10}\rightarrow\ket{...10}, \ket{...11}\rightarrow\ket{...01}.\]
The introduced values of $D^{(k+1)}$ thus form the pattern $a,b,b,a$. 

Splicing the patterns above together, we yield $W^{(k+1)}$, which completes the proof.

\end{proof}


\section{Proof for the Complexity of CompShadow Readout Algorithms}\label{sm:sec:proofs}

\subsection{Complexity of population estimation algorithm}

Theorem~2 asserts the complexity of the CompShadow population estimation algorithm. Before presenting its proof, we first show two lemmas that will assist in the proof. We directly use the lemma in the deductions.
\begin{lemma}\label{lemma:W}
1) Given a Walsh matrix $W$ of size $2^n\times 2^n$, 
\begin{align}
(W^{-1})^2_{ij}&=\begin{cases}
(2^{1-n}-1)^2, &i=j=0,\\
2^{2-2n}, &\text{else},
\end{cases}\\
\|(W^{-1})_{i,*}\|^2&=\begin{cases}
1, &i=0,\\
2^{2-n}, &\text{else},
\end{cases}\\
\|W^{-1}\|^2_F&=5-2^{2-n}.
\end{align}
2) For any state $\rho$ and its population vector $\vec{p}$, suppose the $\ket{0}$ probabilities of its shadows are $\vec{A}$, 
\begin{equation}
\sum_j A_j(1-A_j)=2^{n-2}(1-\sum_a p_a^2).
\end{equation}
\end{lemma}

\begin{proof}[Proof of Lemma~\ref{lemma:W}]
1) It suffices to check $WW^{-1}=I$.

2) It suffices to substitute $\vec{A}=W^{-1}\vec{p}$ (Eqn.~(19) in the maintext) and Equation~(S4) into $\sum_j A_j(1-A_j)$.
\end{proof}

\begin{theorem}
    The CompShadow population estimation method necessitates $O(\frac{1}{\epsilon^{2}}\log(\frac{1}{\eta}))$ shadows and $O(\frac{1}{\epsilon^{4}}\log(\frac{1}{\eta}))$ measurement shots to approximate the population $p_a$ for any bitstring $\ket{a}$ within an error margin of $\epsilon$ and a success probability of $1-\eta$.
    \end{theorem}

\begin{proof}[Proof of Theorem~2]
As described in the maintext, given an arbitrary bitstring $a$, CompShadow readout samples a subvector of $(W^{-1})_{a,*}$, and then conduct measurements for the corresponding shadows. Suppose each corresponding shadow is measured for $N$ shots. The overall error $E_\text{total}$ of estimating $p_a$ includes the statistical error in the shadow measurement and the error of estimating $p_a$ from the subvector sampling method. Mathematically, we have
\begin{equation}
E_\text{total}= E_\text{measurement} +E_\text{sampling}.
\end{equation}

Due to the randomness in each measurement and the sampling, the overall error of the method will fluctuate randomly. We therefore evaluate a upper bound for the expectation of $E_\text{total}$. For the measurement error, the variance of measured $\ket{0}$-probability of the $j$-th shadow is
\begin{equation}
\mathbb{E}(A_j-\tilde{A_j})^2=\frac{A_j(1-A_j)}{N},
\end{equation}
in which $\tilde{A_j}$ is the measured $\ket{0}$-probability of the $j$-th shadow by $N$ shots for $j>0$. $\tilde{A_0}=1$. Then we estimate the measurement error by the Root Mean Square Error (RMSE), as
\begin{align}
E_\text{measurement}&\approx\sqrt{\mathbb{E}(p_a-\tilde{p_a})^2}\nonumber\\
&=\sqrt{\mathbb{E}\left((W^{-1})_{a,*}\;,\;\vec{A}-\tilde{\vec{A}}\right)^2}\nonumber\\
&=\sqrt{\mathbb{E}\left(\sum_{j>0} (W^{-1})_{a,j}(A_j-\tilde{A_j})\right)^2}\nonumber\\
&\leq \sqrt{\sum_{j>0} (W^{-1})_{a,j}^2\cdot\mathbb{E}\sum_{j>0}(A_j-\tilde{A_j})^2}\nonumber\\
&= \sqrt{2^{2-n}\sum\frac{A_j(1-A_j)}{N}}\nonumber\\
&= \sqrt{\frac{1}{N}(1-\sum_a p_a^2)}\nonumber\\
&\leq \sqrt{\frac{1}{N}}.\label{eqn:Emeas}
\end{align}

The IPE algorithm (Alg.~1 in the maintext) guarantee a relative error bound $\xi$ and success rate $1-\eta$ by $O(\frac{\log(1/\eta)}{\xi^2})$ round of sampling $(W^{-1})_{a,*}$ and querying the elements of $\vec{A}$. For a overall error $E_\text{total}\leq \epsilon$, we set $E_\text{measurement}=E_\text{sampling}=\frac{\epsilon}{2}$. For the measurements, by Eqn.~(\ref{eqn:Emeas}), it suffices to take $N\geq\frac{1}{\epsilon^2}$. For the sampling, taking relative error $\xi=\frac{\epsilon}{2}\leq \frac{\epsilon}{2p_a}$ can guarantee $E_\text{sampling}=\frac{\epsilon}{2}$. The number of rounds is then less than $O(\frac{\log(1/\eta)}{\epsilon^2})$. Each round takes $N$ shots to query the elements of $\vec{A}$. The overall number of shots is then $O(\frac{1}{\epsilon^4}\log(\frac{1}{\eta}))$.
\end{proof}

\subsection{Complexity of observable estimation algorithm}

Theorem~3 asserts the complexity of the CompShadow observable estimation algorithm.

\begin{theorem}
    The CompShadow expectation estimation method requires $O(\frac{1}{\epsilon^2})$ copies of the state to estimate the expectation of any Pauli observable within an error of $\epsilon$.
    \end{theorem}

\begin{proof}[Proof of Theorem~3]
From Eqn.~(\ref{eqn:Zrho}) in the maintext, the expectation of $\braket{Z^{\otimes n}}$ under $\rho$ equals the expectation of $\braket{Z}$ under the $2^n-1$-shadow $\text{Tr}_{\{2,...,n\}}(U_{2^n-1}\rho U_{2^n-1}^\dagger)$. For CompShadow readout, the expectation estimator is $V=\frac{\sum V_i}{N}$ such that
\begin{align}
\mathbb{P}(V_i=1)&=\text{Tr}(\ket{0}\bra{0}\otimes I^{\otimes n-1} U_{2^n-1}\rho U_{2^n-1}^\dagger)\nonumber\\
\mathbb{P}(V_i=-1)&=\text{Tr}(\ket{1}\bra{1}\otimes I^{\otimes n-1} U_{2^n-1}\rho U_{2^n-1}^\dagger)\nonumber.
\end{align}

Suppose $N$ measurement shots, the RMSE is
\begin{align}
E_{\text{RMSE}}&=\sqrt{\mathbb{E}(V-\braket{Z})^2}\\
&=\sqrt{\mathbb{E}V^2-\braket{Z}^2}\\
&=\sqrt{\frac{N+(N^2-N)\braket{Z}^2}{N^2}-\braket{Z}^2}\\
&=\sqrt{\frac{1-\braket{Z}^2}{N}}.
\end{align}

To have $E_{\text{RMSE}}\approx \epsilon$, the number of shots $N$ is then $O(\frac{1}{\epsilon^2})$. For direct readout, to evaluate $\braket{Z^{\otimes n}}$, the estimator is $V'=\frac{\sum V'_i}{N}$ such that
\begin{align}
\mathbb{P}(V'=1)&=\sum_{h(i) \equiv 0 (\text{mod} 2)}\text{Tr}(\ket{i}\bra{i}\rho)\nonumber\\
\mathbb{P}(V'=-1)&=\sum_{h(i) \equiv 1 (\text{mod} 2)}\text{Tr}(\ket{i}\bra{i}\rho)\nonumber.
\end{align}

Suppose $N$ measurement shots, the RMSE is
\begin{align}
E'_{\text{RMSE}}&=\sqrt{\mathbb{E}(V'-\braket{Z^{\otimes n}})^2}\\
&=\sqrt{\mathbb{E}V^{\prime 2}-\braket{Z^{\otimes n}}^2}\\
&=\sqrt{\frac{N+(N^2-N)\braket{Z^{\otimes n}}^2}{N^2}-\braket{Z^{\otimes n}}^2}\\
&=\sqrt{\frac{1-\braket{Z^{\otimes n}}^2}{N}}.
\end{align}

To have $E'_{\text{RMSE}}\approx \epsilon$, the number of shots $N$ is then $O(\frac{1}{\epsilon^2})$, which is equal to the requirement of CompShadow readout.

\end{proof}

\begin{figure*}[t]
\begin{center}
\includegraphics[width=0.7\linewidth]{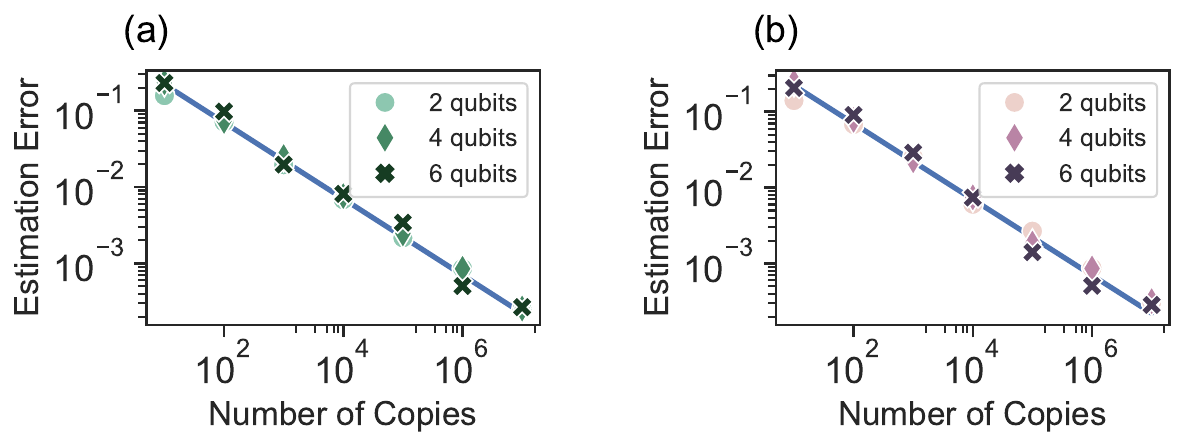}
\end{center}
\caption{\textbf{Numerical validation for the $N=O(1/\epsilon^2)$ scaling of CompShadow and direct readout.} We test the two methods reading the $Z^{\otimes n}$ expectation of random Haar states with varying number of qubits and state copies. (a) The results for CompShadow. (b) The results for direct readout. The blue lines illustrates a $\epsilon=0.7/\sqrt{N}$ function.
}
\label{fig:obsscale}
\end{figure*}

We provide a numerical experiment that validates the number of copies scaling asserted in Theorem~3, as shown in Fig.~\ref{fig:obsscale}. We vary the system size and the number of copies for CompShadow and direct readout reading the $Z^{\otimes n}$ expectation of random Haar states. We find the two methods achieves very close estimation error with same number of copies, as the data points are all approximately distributed along the same blue line $\epsilon=0.7/\sqrt{N}$. 

\section{Supporting Data for Numerical Experiments}\label{sm:sec:more_sim}

We introduce the noise models for gate errors and correlated measurement errors in the numerical experiments, as well as validation data for the choice of algorithm parameters in the experiments.
\subsection{Noise model}

\begin{figure*}[t]
\begin{center}
\includegraphics[width=\linewidth]{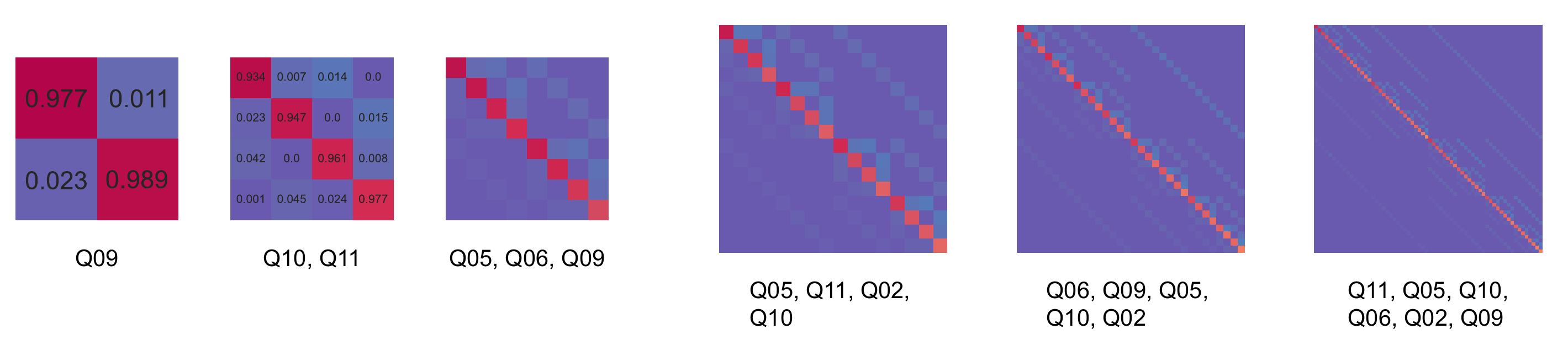}
\end{center}
\caption{\textbf{The probablistic transition matrices during measurements on six qubits of \textit{Zuchongzhi} 2.1.} Due to the complexities of the correlations among qubits, for each combination of these qubits, there is a unique measurement error channel. Here we demonstrate a few samples of the transition matrices with varying system size. The texts below them show the corresponding qubits.
}
\label{fig:Tmat}
\end{figure*}

The single-qubit and two-qubit gate errors are simulated by the commonly-used Pauli noise channels
\begin{align*}
\mathcal{E}_{e_1}(\rho)&=(1-e_1)\rho+\frac{e_1}{3}\sum_i \sigma_i\rho\sigma_i,\\
\mathcal{E}_{e_2}(\rho)&=(1-e_2)\rho+\frac{e_2}{15}\sum_i \sigma_i\rho\sigma_i,
\end{align*}
in which $e_1,e_2$ are the typical gate error rates (single-qubit $e_1=0.16\%$, two-qubit $e_2=0.6\%$, respectively) of \textit{Zuchongzhi} 2.1~\cite{zhu2021quantum}. For the idle windows in the compression circuit, we simulate $T_1$ decay by amplitude damping channel
\begin{align*}
\mathcal{E}_{\text{AD}}(\rho)&=E_0\rho E_0^\dagger+E_1\rho E_1^\dagger,\\
E_0&=\begin{pmatrix}
1&0\\
0&\sqrt{1-\gamma}
\end{pmatrix},\\
E_1&=\begin{pmatrix}
0&\sqrt{\gamma}\\
0&0
\end{pmatrix},
\gamma=1-e^{-T_{\text{gate}}/T_1},
\end{align*}
in which $T_{\text{gate}}=24\text{ns}$ is the iSWAP-like gate duration, $T_1=26.5\mu\text{s}$ is the average relaxation time. The measurement errors are mostly probablistic transitions among basis states during computational basis measurements. We yield matrices of the transition probabilities from the joint calibration data on $\text{Q02},\text{Q05},\text{Q06},\text{Q09},\text{Q10},\text{Q11}$ of \textit{Zuchongzhi} 2.1, as demonstrated in Fig.~\ref{fig:Tmat}. In the simulation, when conducting a $n$-qubit measurement, we randomly select $n$ qubits from the calibrated qubit list, and apply the corresponding transition matrix $T$ to the amplitude population vector, as $\vec{p}_{\text{readed}}=T\vec{p}_{\text{prior}}$. 
In the simulations of tensor product noise inversion (TPN) and unfolding methods, we also utilize these calibrated single-qubit transition matrices to conduct noise inversion on the direct measurement results, instead of recalibrating the simulated measurement errors. This approach ignores the shot noise introduced by the finite measurement shots of calibration and focuses primarily on the methods' ability to perform error mitigation given the perfect calibration data.

\subsection{The parameters setting in numerical simulations}

\begin{figure*}[t]
\begin{center}
\includegraphics[width=0.75\linewidth]{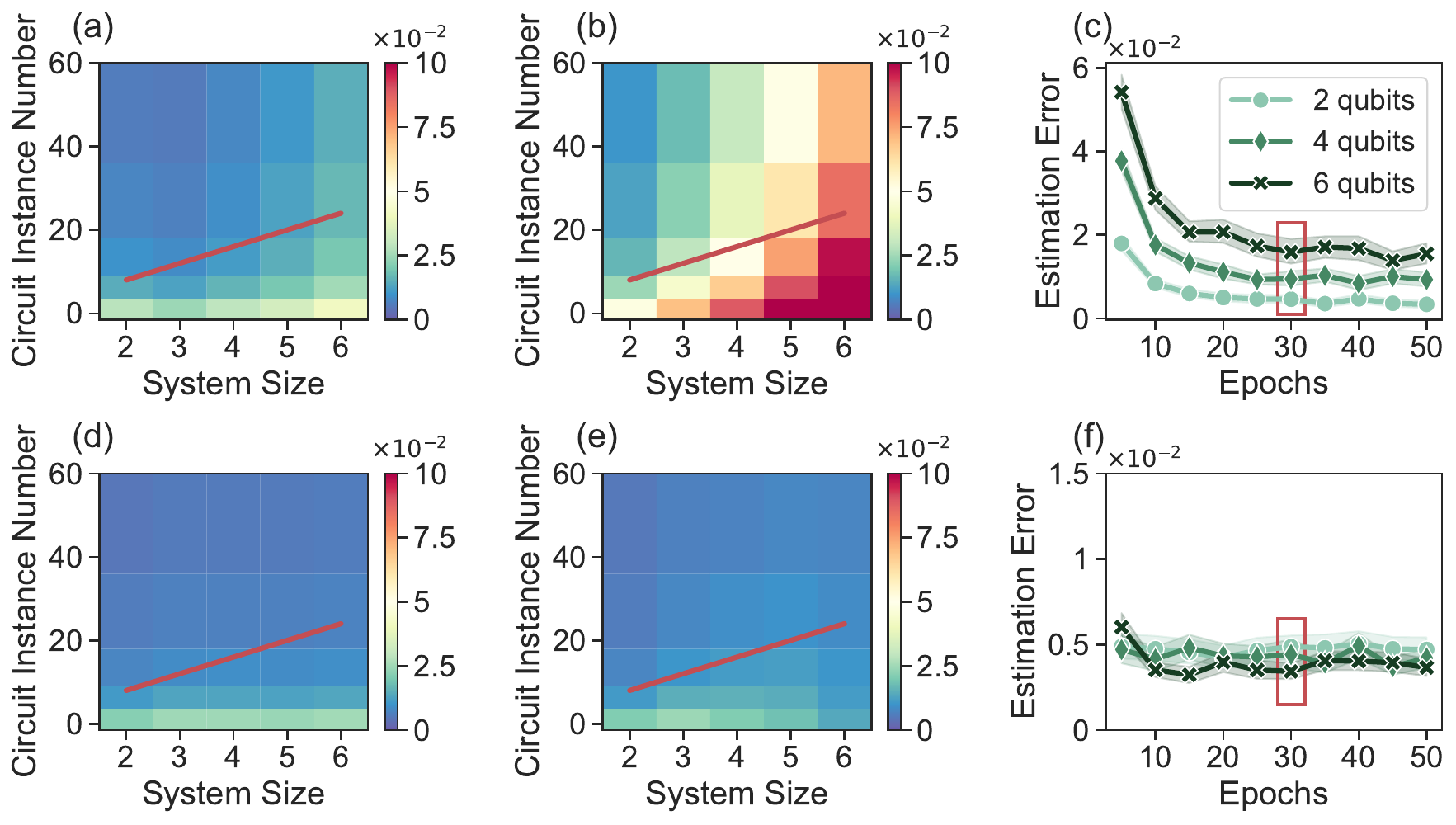}
\end{center}
\caption{\textbf{Validation of parameters setting for CompShadow, model-free and unfolding methods.} The two rows show the estimation errors of these methods reading random basis states and random Haar states, respectively. (a,d) Estimation error versus system size and number of circuit instances of CompShadow. The red lines mark our choice of the number of circuit instances as $4n$. (b,e) Estimation error versus system size and circuit instance number of the model-free method. The red lines mark our choice of the circuit instance number as $4n$. (c,f) Estimation error versus system size and epoch number of the unfolding method. The red rectangle mark our choice of the epoch number as $30$. All subplots depict average results obtained from 100 independent repeated experiments. Uncertainty bands around the curves represent the standard errors of the means.
}
\label{fig:sm-para}
\end{figure*}

The algorithm parameters among the simulated methods include the number of circuit instances in randomized compiling (RC) and model-free method, and the number of iteration epochs of unfolding method. In the maintext, they are taken as $4n$ and $30$ respectively ($n$ is system size.). Here, we show the validity such choices through numerical experiments. As shown in Fig.~\ref{fig:sm-para}, we vary the parameters of the algorithms with increasing system size, reading the $Z^{\otimes n}$ expectation for states sampled from the set of $n$-qubit basis states, and the Haar distribution, respectively. The performance of all three methods deteriorates as the system size increases, but increasing the parameters for model-free and unfolding methods helps improve their performance. In contrast, the impact of number of circuit instances on the performance of CompShadow is less. For the model-free method, in order to maintain a high performance, it is necessary to grow the circuit instance number with the system size, thus we chose the value of $4n$, shown by the red line in Fig.~\ref{fig:sm-para}b,e. In order to maintain consistent number of shots between CompShadow and the model-free method, we let the number of circuit instances of CompShadow equals $4n$ as well. The performance of the unfolding method remains stable when the number of epochs is increased to 30, hence we chose the value 30 in our main experiment, as indicated by the rectangle in Fig.~\ref{fig:sm-para}c,f. 

\section{CompShadow Probing Method for Local Density}\label{sm:sec:local_density}

The local density\cite{cramer2008exploring,zhu2022observation} of a many-body system is defined as the expectation of observable
\begin{align}
\hat{n}_{\ket{0}}&=\frac{1}{N_0}\sum_{m=1}^{N_0}\hat{\sigma}^+_m\hat{\sigma}^-_m\\
&=\frac{1}{N_0}\sum_{m=1}^{N_0}(\ket{0}\bra{0})_m,\nonumber
\end{align}
in which $N_0$ is the size of a subsystem. Essentially, we need to find the $\ket{0}$ probabilities of each qubit in the system.

Suppose the system has $n$ qubits, the $\ket{0}\bra{0}$ observables of the probed qubits are all diagonal, as 
\begin{align*}
\text{Diag}(O_1)&=(1,0)\otimes(1,1)^{\otimes n-1}\\
\text{Diag}(O_2)&=(1,1)\otimes (1,0)\otimes(1,1)^{\otimes n-2}\\
...\\
\text{Diag}(O_n)&=(1,1)^{\otimes n-1}\otimes(1,0).
\end{align*}

Same as the deduction in Eqn.~(8), we have
\begin{equation}
\braket{O_i}=(\vec{p},\text{Diag}(O_i))=A_{2^{n-i}},
\end{equation}
which are exactly ones of the compression shadows. Therefore, we can estimate $\braket{O_i}$ by measuring the $\ket{0}$ probability of the $2^{n-i}$-th shadow of $\rho$.

\section{Compression Scheduling}\label{sm:sec:schedule}

In the maintext, the CompShadow methods measure the first qubit of the whole register. In this note, we introduce how to redesign the compression circuit to move the measurement location, and how to design constant-depth compression circuits by accepting more measured qubits, on the task of observable estimation.

\subsection{Methods}
\begin{figure*}[t]
    \begin{center}
    \includegraphics[width=1\linewidth]{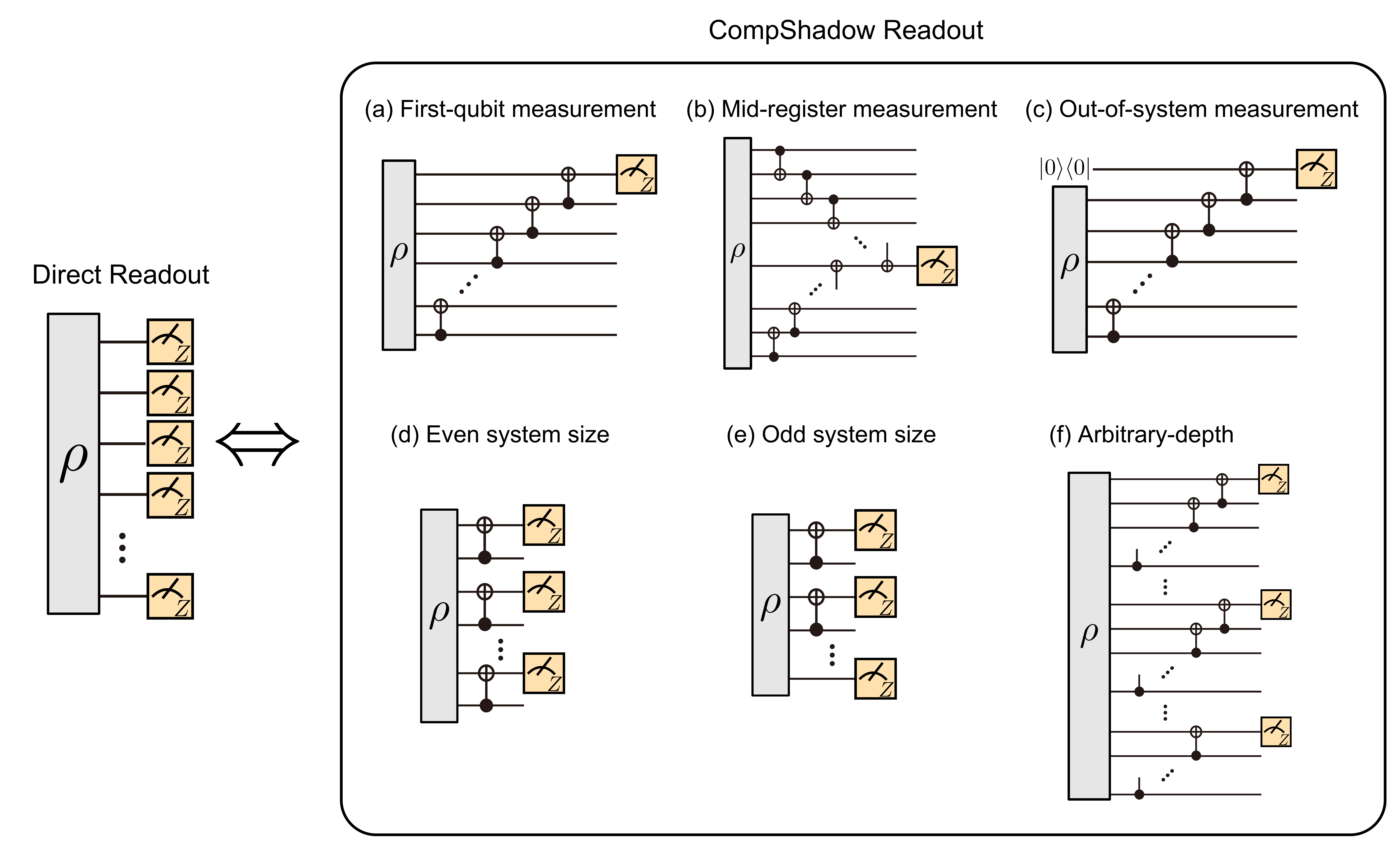}
    \end{center}
    \caption{\textbf{Compression Scheduling methods for observable expectation estimation.} The direct measurements on all qubits of $\rho$ can be equivalently conducted by the circuits a-f. (a) The $n-1$-depth CompShadow that only require first-qubit measurements. (b) The $\text{max}(k,n-k+1)$-depth CompShadow that require measurements on the $k$-qubit. (c) The CompShadow that conduct measurements out of the $\rho$ system. (d,e) The one-depth CompShadow for even and odd system size, respectively. (f) Arbitrary-depth CompShadow, which can be viewed as an interpolation of (a) and (d,e).}
    \label{fig:scheduling}
    \end{figure*}

\begin{figure*}[t]
\begin{center}
\includegraphics[width=0.6\linewidth]{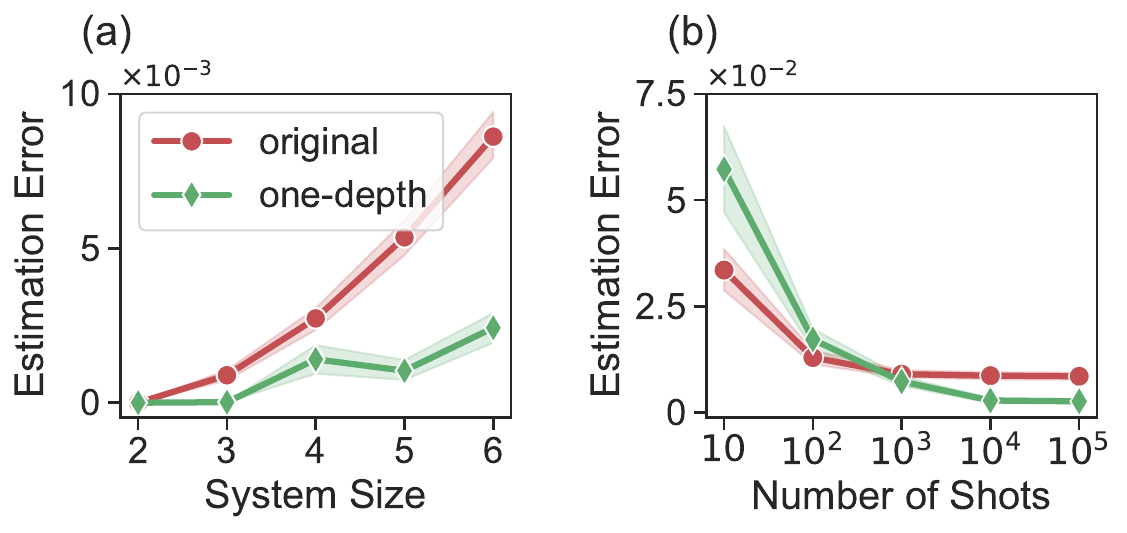}
\end{center}
\caption{\textbf{Comparative error analysis of the original CompShadow and one-depth CompShadow methods.} (a) Error curves of the two methods with varying system size, given infinite measurement shots. (b) Error curves of the two methods with varying number of measurement shots, reading six-qubit random basis states. All subplots depict average results obtained from 100 independent repeated experiments. Uncertainty bands around the curves represent the standard errors of the means. These subfigures generally show that one-depth CompShadow can adapt to the noise characteristics of 2-local correlated measurement errors and hence performs better in such case.
}
\label{fig:plt-half-comp}
\end{figure*}

Without loss of generality, we assume the observable to estimate is $Z^{\otimes n}$. Originally, the compression circuit is
\begin{align}
    U=\prod^{1}_{i=n-1}\text{CNOT}(i+1,i),\label{eqn:sm-ulast}
\end{align}
which transform the observable to 
$Z\otimes I^{\otimes n-1}$, as
\begin{align}
\tr{\rho Z^{\otimes n}}=\tr{U\rho U^\dagger Z\otimes I^{\otimes n-1}}.\label{eqn:Zrho}
\end{align}

Therefore it suffices to measure the first qubit.
To move the measurement qubit to the $k$-th qubit, we let
\begin{align}
U_k=\prod^{k}_{i=n-1}\text{CNOT}(i+1,i)\prod^{k-1}_{i=1}\text{CNOT}(i,i+1),
\end{align}
then
\begin{align}
\tr{\rho Z^{\otimes n}}=\tr{U_k\rho U_k^\dagger I^{\otimes k-1}\otimes Z\otimes I^{\otimes n-k}},
\end{align}
and the measurement location is effectively moved, as shown in Fig.~\ref{fig:scheduling}b.

To move the measurement qubit outside the original system, we use an ancilla qubit of state $\ket{0}$. By applying the compression circuit on the composite system, we yield the same expectation $\tr{\ket{0}\bra{0}\otimes \rho Z^{\otimes n+1}}=\tr{\rho Z^{\otimes n}}$, as as shown in Fig.~\ref{fig:scheduling}c.

Besides, applying $V=\prod^{\ceil{\frac{n-1}{2}}}_{i=1}\text{CNOT}(2i,2i-1)$ on $\rho$, we have
\begin{align}
    \tr{\rho Z^{\otimes n}}=
    \begin{cases}
        \tr{V\rho V^\dagger (Z\otimes I)^{\otimes \frac{n}{2}}},&\  n\text{ even}\\
        \tr{V\rho V^\dagger (Z\otimes I)^{\otimes \frac{n-1}{2}}\otimes Z},&\  n\text{ odd}
    \end{cases}
\end{align}
which allows us to estimate $\tr{\rho Z^{\otimes n}}$ by measuring half of the $n$ qubits, while the compression circuit $V$ have only depth 1. Generally, we can design compression circuit of depth $l$ that requires about $\frac{n}{l+1}$ measurement qubits to conduct the observable estimation, as drawn in Fig.~\ref{fig:scheduling}f. 

\subsection{Numerical demonstration for one-depth CompShadow}

The compression scheduling techniques enables us to flexibly design the compression circuits and measurement locations, thereby can adapt to the specific defects and noise characteristics of quantum processors in state readout. For instance, the CompShadow readout by one-depth circuit (coined ``one-depth CompShadow'') in Fig.~\ref{fig:scheduling}d,e can avoid the 2-local correlated measurement errors while minize the effect of gate errors and thermal relaxations. A numerical validation is shown in Fig.~\ref{fig:plt-half-comp}. We first construct a 2-local correlated measurement error channel by taking tensor product of 1 or 2-qubit calibrated probablistic transition matrices shown in Fig.~\ref{fig:Tmat}. Then we test both the original CompShadow and the one-depth CompShadow in the task of finding the $Z^{\otimes n}$ expectation of random basis states. Both methods are helped by RC with number of circuit instances of $4n$ and $16$, respectively. We find the one-depth CompShadow generally performs better than the original CompShadow, while using much less number of circuit instances. Besides, the estimation errors of original CompShadow under the 2-local measurement errors are similar to the ones under full measurement errors displayed in Fig.~2 in the maintext, indicating the correlated measurement errors observed in \textit{Zuchongzhi}~2.1 is primarily 2-local.

\bibliographystyle{apsrev4-1}

\bibliography{b}